\documentclass[twocolumn]{aastex63}
\usepackage{graphicx}
\usepackage{ulem}
\usepackage{cellspace}
\usepackage{amssymb}
\usepackage{amsmath}
\usepackage{longtable}
\usepackage{tabu}
\usepackage{color}
\usepackage{etoolbox}
\usepackage{supertabular}
\usepackage{dcolumn}
\usepackage{ wasysym }

\newcommand\Tstrut{\rule{0pt}{2.9ex}}       
\newcommand\Bstrut{\rule[-1.3ex]{0pt}{0pt}} 
\newcommand\TBstrut{\Tstrut\Bstrut}

\submitjournal{ApJ}
\accepted{May 11, 2020}

\def\Tcoh{T_{\textrm{\mbox{\tiny{coh}}}}}
\def\Tref{T_{\textrm{\mbox{\tiny{ref}}}}}

\def\EatH{Einstein@Home}

\def\sci#1#2{#1\times10^{#2}}

\newcommand{\avgSeg}[1]{\overline{#1}}			

\newcommand{\Freq}{f}
\newcommand{\fdot}{{\dot{\Freq}}}
\newcommand{\fddot}{\ddot{\Freq}}

\newcommand{\M}{\mathcal{M}}

\newcommand{\Gauss}{\mathrm{\MakeUppercase{G}}}
\newcommand{\Signal}{{\mathrm{\MakeUppercase{S}}}}
\newcommand{\Line}{{\mathrm{\MakeUppercase{L}}}}
\newcommand{\Transient}{{\mathrm{t\MakeUppercase{L}}}}

\newcommand{\NoisetL}{{\Gauss\Line\Transient}}

\providecommand{\sc}[1]{\widehat{#1}}
\renewcommand{\sc}[1]{\widehat{#1}}

\newcommand{\BSNtsc}{{\hat\beta}_{{\Signal/\NoisetL}}}	

\newcommand{\BSGLtLr}{{\hat\beta}_{{\Signal/\NoisetL r}}}

\newcommand{\F}{\mathcal{F}}		

\newcommand{\avF}{\avgSeg{\F}}

\newcommand{\Nseg}{{N_{\mathrm{seg}}}}

\newcommand{\G}{1WGA J1713.4-3949}
\newcommand{\V}{CXOU J085201.4-461753}
\newcommand{\C}{CXOU J232327.9+584842}

\newcommand{\Gshort}{J1713}
\newcommand{\Vshort}{J085201}
\newcommand{\Cshort}{J232327}

\newcommand{\posCasAa}{\ensuremath{6.123770}}
\newcommand{\posVelaJra}{\ensuremath{2.3213891}}
\newcommand{\posGa}{\ensuremath{4.5093705}}
\newcommand{\posCasAd}{\ensuremath{1.026458}}
\newcommand{\posVelaJrd}{\ensuremath{-0.8080543}}
\newcommand{\posGd}{\ensuremath{-0.6951891}}

\newcommand{\paramfdothiCasA}{\ensuremath{0}} 
\newcommand{\paramfdotloCasA}{\ensuremath{-\sci{9.6}{-9}}}
\newcommand{\paramfdotloVela}{\ensuremath{-\sci{4.5}{-9}}} 
\newcommand{\paramfdothiVela}{\ensuremath{0}}
\newcommand{\paramfdothiG}{\ensuremath{0}} 
\newcommand{\paramfdotloG}{\ensuremath{-\sci{2.0}{-9}}} 
\newcommand{\paramfddothiCasA}{\ensuremath{\sci{4.6}{-18}}} 
\newcommand{\paramfddotloCasA}{\ensuremath{0}}
\newcommand{\paramfddotloVela}{\ensuremath{0}} 
\newcommand{\paramfddothiVela}{\ensuremath{\sci{1.0}{-18}}}
\newcommand{\paramfddothiG}{\ensuremath{\sci{2.0}{-19}}}
\newcommand{\paramfddotloG}{\ensuremath{0}} 

\newcommand{\ageCasA}{\ensuremath{330} }
\newcommand{\ageVela}{\ensuremath{700} }
\newcommand{\ageG}{\ensuremath{1600}}

\newcommand{\distanceC}{\ensuremath{3.4} }
\newcommand{\distanceVnear}{\ensuremath{0.2}}
\newcommand{\distanceVfar}{\ensuremath{1}}
\newcommand{\distanceG}{\ensuremath{1.3}}

\newcommand{\TrefGPS}{\ensuremath{1131943508}}

\newcommand{\ThrVelalow}{3.10}
\newcommand{\ThrVelamid}{4.9}
\newcommand{\ThrVelahigh}{5.86}

\newcommand{\ThrCasAlow}{2.75}
\newcommand{\ThrCasAmid}{4.75}
\newcommand{\ThrCasAhigh}{5.42}

\newcommand{\ThrGlow}{2.95}
\newcommand{\ThrGmid}{4.05}
\newcommand{\ThrGhigh}{5.02}

\newcommand{\NsftOOneToOTwoOne}{86}
\newcommand{\NsftOOneToOTwoTwo}{76}

\newcommand{\NsftOTwoTwoToOTwoOne}{89}

\newcommand{\pvalueOTwoOneWOneTempl}{7.6\times{10^{-5}}}

\newcommand{\FreqOutlierHR}{368.8013845}
\newcommand{\fdotOutlierHR}{-4.378054\times 10^{-9}}
\newcommand{\fddotOutlierHR}{5.7\times 10^{-19}}
\newcommand{\TwoFOutlierHR}{67.0}

\newcommand{\TwoFOutlierHRNormScore}{1}

\newcommand{\FreqOutlierOTwoOne}{368.8013794}
\newcommand{\fdotOutlierOTwoOne}{-4.37800\times 10^{-9}}
\newcommand{\fddotOutlierOTwoOne}{5.9\times 10^{-19}}
\newcommand{\TwoFOutlierOTwoOne}{57.0}
\newcommand{\TwoFOutlierOTwoOneNormScore}{0.98}

\newcommand{\FreqOutlierOTwoTwo}{368.8014296}
\newcommand{\fdotOutlierOTwoTwo}{-4.37800\times 10^{-9}}
\newcommand{\fddotOutlierOTwoTwo}{5.5\times 10^{-19}}
\newcommand{\TwoFOutlierOTwoTwo}{34.7}
\newcommand{\TwoFOutlierOTwoTwoNormScore}{0.64} 
\newcommand{\TwoFGNExpectedOTwoTwo}{32.3}
\newcommand{\TwoFOTwoTwopvalue}{10\%}

\newcommand{\FreqOutlierOTwo}{368.8013795}
\newcommand{\fdotOutlierOTwo}{-4.37800\times 10^{-9}}
\newcommand{\fddotOutlierOTwo}{5.9\times 10^{-19}}
\newcommand{\TwoFOutlierOTwo}{53.3}
\newcommand{\TwoFOutlierOTwoNormScore}{0.48} 

\newcommand{\SignalExpectedOOneOTwo}{171.8}
\newcommand{\SignalStdOOneOTwo}{26}
\newcommand{\NtemplOOneOTwo}{2.9\times 10^{10}}
\newcommand{\FreqOutlierOOneOTwo}{368.8013845}
\newcommand{\fdotOutlierOOneOTwo}{-4.37805\times 10^{-9}}
\newcommand{\fddotOutlierOOneOTwo}{5.7\times 10^{-19}}
\newcommand{\TwoFOutlierOOneOTwo}{76.0}
\newcommand{\TwoFOutlierOOneOTwoNormScore}{0.44} 

\newcommand{\NcandVelaInStageOne}{3275}

\newcommand{\NcandCasAInStageOne}{2269}

\newcommand{\NcandGInStageOne}{3918}

\newcommand{\NcandVelaOutStageTwo}{142}
\newcommand{\NcandCasAOutStageTwo}{81}
\newcommand{\NcandGOutStageTwo}{352}

\newcommand{\NcandVelaInStageThree}{137} 
\newcommand{\NcandCasAInStageThree}{78} 
\newcommand{\NcandGInStageThree}{331}
\newcommand{\NcandTotInStageThree}{546}

\newcommand{\NrejectDMoffVela}{557}
\newcommand{\NrejectDMoffCasA}{1223}
\newcommand{\NrejectDMoffG}{1032}

\newcommand{\lowestULc}{$1.2\times 10^{-25}$}

\newcommand{\lowestULv}{$9.3\times 10^{-26}$}
\newcommand{\lowestULvfreq}{172.5}
\newcommand{\lowestULg}{$8.8\times 10^{-26}$}

\shorttitle{Search for CWs from the central compact object in SNRs Cas A, Vela Jr. and G347.3$-$0.5}
\shortauthors{Papa et al.}

\begin{document}

\title{Search for Continuous Gravitational Waves from the Central Compact Objects in Supernova Remnants Cassiopeia A, Vela Jr. and G347.3$-$0.5}

\correspondingauthor{M.Alessandra Papa}
\email{maria.alessandra.papa@aei.mpg.de}

\author[0000-0002-1007-5298]{M. Alessandra Papa}
\affiliation{Max Planck Institute for Gravitational Physics (Albert Einstein Institute), Callinstrasse 38, 30167 Hannover, Germany}
\affiliation{University of Wisconsin Milwaukee, 3135 N Maryland Ave, Milwaukee, WI 53211, USA}
\affiliation{Leibniz Universit\"at Hannover, D-30167 Hannover, Germany}

\author{J. Ming}
\email{jing.ming@aei.mpg.de}
\affiliation{Max Planck Institute for Gravitational Physics (Albert Einstein Institute), Callinstrasse 38, 30167 Hannover, Germany}
\affiliation{Leibniz Universit\"at Hannover, D-30167 Hannover, Germany}

\author{E.V. Gotthelf}
\email{eric@astro.columbia.edu}
\affiliation{Columbia Astrophysics Laboratory, Columbia University, 550West 120th Street, New York, NY 10027-6601, USA}

\author{B. Allen}
\affiliation{Max Planck Institute for Gravitational Physics (Albert Einstein Institute), Callinstrasse 38, 30167 Hannover, Germany}
\affiliation{University of Wisconsin Milwaukee, 3135 N Maryland Ave, Milwaukee, WI 53211, USA}
\affiliation{Leibniz Universit\"at Hannover, D-30167 Hannover, Germany}

\author{R. Prix}
\affiliation{Max Planck Institute for Gravitational Physics (Albert Einstein Institute), Callinstrasse 38, 30167 Hannover, Germany}
\affiliation{Leibniz Universit\"at Hannover, D-30167 Hannover, Germany}

\author{V. Dergachev}
\affiliation{Max Planck Institute for Gravitational Physics (Albert Einstein Institute), Callinstrasse 38, 30167 Hannover, Germany}
\affiliation{Leibniz Universit\"at Hannover, D-30167 Hannover, Germany}

\author{H.-B. Eggenstein}
\affiliation{Max Planck Institute for Gravitational Physics (Albert Einstein Institute), Callinstrasse 38, 30167 Hannover, Germany}
\affiliation{Leibniz Universit\"at Hannover, D-30167 Hannover, Germany}

\author{A. Singh}
\affiliation{Max Planck Institute for Gravitational Physics (Albert Einstein Institute), Callinstrasse 38, 30167 Hannover, Germany}
\affiliation{Leibniz Universit\"at Hannover, D-30167 Hannover, Germany}
\affiliation{The Geophysical Institute, Bjerknes Centre for Climate Research, University of Bergen, 5007 Bergen, Norway}

\author{S. J. Zhu}
\affiliation{Max Planck Institute for Gravitational Physics (Albert Einstein Institute), Callinstrasse 38, 30167 Hannover, Germany}
\affiliation{DESY, 15738 Zeuthen, Germany}

\begin{abstract}
  We perform a sub-threshold follow-up search for continuous nearly-monochromatic 
  gravitational waves from the central compact
  objects associated with the supernova remnants Vela Jr.,
  Cassiopeia A, and SNR G347.3$-$0.5.  
  Across the three targets, we investigate the most promising $\approx$10,000 combinations of gravitational wave frequency and frequency derivative values, based on the results from an \EatH\ search of the LIGO O1 observing run data, dedicated to these objects. The selection threshold is set so that a signal could be confirmed using the newly released O2 run LIGO data. In order to achieve best sensitivity we perform two separate follow-up searches, on two distinct stretches of the O2 data. Only one candidate survives the first O2 follow-up investigation, associated with the central compact
  object in SNR G347.3$-$0.5, but it is not conclusively confirmed.
  In order to assess a possible astrophysical origin we use archival X-ray observations and search for amplitude modulations of a pulsed signal at the putative rotation frequency of the neutron star and its harmonics.
  This is the first extensive electromagnetic follow-up of a
  continuous gravitational wave candidate performed to date. No significant
  associated signal is identified. New X-ray observations contemporaneous with the LIGO O3 run will enable a more 
sensitive search for an electromagnetic counterpart. A focused gravitational wave search in O3 data based on the parameters provided here should be easily able to shed light on the nature of this outlier. Noise investigations on the LIGO instruments could also reveal the presence of a coherent contamination.

\end{abstract}

\keywords{gravitational waves --- continuous --- SNRs --- pulsar -- J232327.9+584842 --- CXOU J085201.4-461753 --- 1WGA J1713.4-3949}

\section{Introduction}
\label{sec:introduction}

At the time of writing, gravitational waves from 14 binary black hole mergers and a binary neutron star merger have been 
reported \citep{Abbott:2020uma,Venumadhav:2019lyq,Nitz:2019hdf,LIGOScientific:2018mvr}. All of these signals are visible in the detectors for less than a second, or several tens of seconds for a binary neutron star merger, and present a strain amplitude of $\sim 10^{-21}$ for most of that time. 

This paper is about a different type of signal: a persistent train of nearly monochromatic gravitational waves with an amplitude four (or more) orders of magnitude weaker than that of the binary merger signals.

Various physical mechanisms have been suggested that could give rise to continuous gravitational waves:
deformations of the compact object, including asymmetric mass distribution \citep{McDanielJohnsonOwen}, ringing (r-modes) \citep{Owen:1998xg}, and  non axi-symmetric flows of bulk matter induced by differential rotation between the core of the star and its crust through Ekman pumping \citep{Singh:2016ilt}, as well as more exotic scenarios, for instance involving very low mass cold dark matter binaries \citep{Horowitz:2019aim,Horowitz:2019pru} or emission from boson clouds around black holes \citep{Arvanitaki+2015,Zhu:2020tht}. 

The relationship between the spin frequency of the star  and the frequency of the continuous gravitational wave signal depends on the emission mechanism but typically the continuous gravitational wave frequency does not exceed twice the spin frequency. So, in the high sensitive band of the LIGO detectors -- between 20 Hz and 2 kHz -- we expect continuous gravitational waves from neutron stars with spin periods in the range 1-100 ms.

Perhaps among the most promising candidates for continuous
gravitational wave emission are young, isolated neutron stars. These could have a high spindown and hence a potentially large energy budget to be radiated away in gravitational waves. 

The central compact objects associated with supernova remnants could indeed be young isolated neutron stars.
We know of $\sim$ 10 objects at the center of supernova remnants \citep{Gotthelf:2013sa}.
Three of these objects are known pulsars (with spin periods $P= 105,112,424$~ms) whose
slow spin-down rate implies a low spin-down power and weak magnetic
dipole field. The rest are undetected as pulsars and their timing
properties remain a mystery. 

In the past the youngest/closest supernova remnants with identified central compact objects have been chosen as targets for   
gravitational wave searches that span a broad range of physically plausible signal frequencies and
frequency-derivatives \citep{Lindblom:2020rug,Millhouse:2020jlt,Ming:2019xse,Abbott:2018qee,Zhu:2016ghk,Aasi:2014ksa}.

In this paper we use the released LIGO O2 data set to further the results of the O1-data search \citep{Ming:2019xse} for continuous gravitational wave emission from the compact object in the supernova remnants Vela Jr., Cassiopeia A and SNR G347.3$-$0.5. The targets and search setups were chosen based on an optimization scheme that maximizes the total detection probability over search setups and astrophysical targets, assuming a total computing budget of a few months on the \EatH\ volunteer distributed computing project \citep{EatH,Ming:2015jla,Ming:2017anf}. The detection probability depends on the assumed source and signal properties (distance, age, ellipticity, frequency) as well as on the sensitivity of the search pipeline (search setup and search parameters, noise level).

In Section~\ref{subsec:o1stage0}, we present the results of the sub-threshold search of 
\EatH\ candidates from \cite{Ming:2019xse}.  In Section~\ref{subsec:o21search} 
we describe the follow-up search using part of the LIGO O2 data set. With this search we
find a marginal candidate for the source associated with SNR G347.3$-$0.5, which is however not definitely confirmed in 
the second part of the O2 data (Section~\ref{sec:g347CandFUs}).
We report a search for an electromagnetic counterpart for this candidate in Section~\ref{sec:g347CandEM}. In Section~\ref{sec:upperlimits}
we compute the upper limits on the intrinsic gravitational wave amplitude stemming from the LIGO search. Finally, in
Section~\ref{sec:conclusions}, we summarize our conclusions.

\begin{deluxetable*}{lccc}
\tablecaption{Search ranges. The spindown ranges quoted are the ones used at $f=$ 100 Hz. The ranges at different frequencies are readily derived from Eq.~\ref{eq:Priors}. The barycentric reference time (the epoch of the ephemeris) is $\Tref=\TrefGPS$ GPS s corresponding to MJD 57345.1986016667 (TDB). \label{tab:SearchParams}
}
\tablehead{
\colhead{COMPACT OBJECT}  & \colhead{\Cshort} & \colhead{\Vshort} & \colhead{\Gshort}\\
\colhead{REMNANT}  & \colhead{(G111.7-02.1 or Cas A)} & \colhead{(G266.2-1.2 or Vela Jr.)} & \colhead{(G347.3-0.5)}\\
  \colhead{\small{references}} & \colhead{\citep{Tananbaum1999}} & \colhead{\citep{Pavlov_2001}} & \colhead{\citep{Mignani:2008ew}}
 }
\startdata
 \TBstrut$f$ range &  {[20-1500] Hz} &  {[20-1500] Hz} &  {[20-1500] Hz} \\
\TBstrut$\fdot $ range (@ 100 Hz) &  [\paramfdotloCasA - \paramfdothiCasA~ ] Hz/s & [\paramfdotloVela - \paramfdothiVela~ ]  Hz/s & [\paramfdotloG - \paramfdothiG ~] Hz/s \\
 \TBstrut$\fddot $ range (@ 100 Hz) &  [ \paramfddotloCasA~  -  \paramfddothiCasA~ ] Hz/s$^2$  & [ \paramfddotloVela~ - \paramfddothiVela ~] Hz/s$^2$   & [ \paramfddotloG~ - \paramfddothiG~ ] Hz/s$^2$  \\
\TBstrut  RA (J2000, in radians) & \posCasAa & \posVelaJra & \posGa \\  
\TBstrut DEC  (J2000, in radians)   &  \posCasAd & \posVelaJrd & \posGd \\
\enddata
\end{deluxetable*}

\begin{deluxetable*}{cccc}
\tablecaption{Spacings on the signal parameters used for the templates in the \EatH~ search. \label{tab:GridSpacings}
}
\tablehead{
 \colhead{COMPACT OBJECT}  & \colhead{\Cshort} & \colhead{\Vshort} & \colhead{\Gshort}\\
 \colhead{REMNANT} & \colhead{G111.7-02.1 or Cas A} & \colhead{G266.2-1.2 or Vela Jr.} & \colhead{G347.3-0.5}
 }
\startdata 
\TBstrut$\Tcoh$ & 245 hr & 369 hr& 489 hr\\
 \hline
\TBstrut$\Nseg$ & 12 & 8 & 6 \\
\TBstrut$\delta f$ & $6.85 \times 10^{-7}$ Hz & $3.21 \times 10^{-7}$ Hz & $2.43 \times 10^{-7}$ Hz \\
 \TBstrut$\delta {\dot{f_c}}$ &  $3.88\times 10^{-12}$ Hz/s  & $1.33\times 10^{-12}$ Hz/s  & $6.16 \times 10^{-13}$ Hz/s \\
\TBstrut$\gamma_1$ & 5 &  9 & 9 \\
\TBstrut$\delta {\ddot{f_c}}$ & $ 4.03\times 10^{-18} {\textrm{Hz/s}}^2 $ & $1.18\times 10^{-18} {\textrm{Hz/s}}^2 $ & $5.07 \times 10^{-19} {\textrm{Hz/s}}^2$ \\
\TBstrut$\gamma_2$ & 21 & 21 & 11 \\
\enddata
\end{deluxetable*}

\section{The Search}
\label{sec:search}

The search described in this paper targets nearly monochromatic
gravitational waves (see for example Section II of
\cite{Jaranowski:1998qm}) from compact objects in three supernova remnants,
\C\ in Cassiopeia A \citep{Tananbaum1999}, \V\ in Vela Jr. \citep{Slane:2000gr},
 and \G\ in SNR G347.3$-$0.5 \citep{pfeffermann1996}, referred to
 herein as \Cshort, \Vshort, and \Gshort, respectively.  No pulsations have been reported from any of these sources.

We use public data from the first (O1; 2015-2016) and second (O2; 2017) run of the
advanced LIGO detectors  \citep{TheLIGOScientific:2016agk,TheLIGOScientific:2014jea,O1data,O2data,Vallisneri:2014vxa}. Measured on a fully coherent search, and assuming that the O1 and O2 sensitivities are comparable, the relative strain sensitivity improvement from using the O2 data over the O1 data is $\sim$ 27\%. But the real value of the O2 data for follow-ups of candidates, is {{\it not}} in the raw sensitivity gain. It rather lies in that the new data is independent of the data where the candidates are originally identified. Together with the fact that the follow-up searches cover a more limited parameter space than the original ones, this makes it possible to assess the significance of a finding on the new data, including confidently claiming a detection.

We use a large-scale hierarchical scheme that was first deployed in \cite{Papa:2016cwb} and that has now become standard practice in broad continuous wave searches \citep{Abbott:2017pqa,Dergachev:2019pgs,Pisarski:2019vxw,Dergachev:2019wqa,Palomba:2019vxe,Dergachev:2019oyu}. The first, and most computationally intensive  search on the O1 data is carried out on the \EatH~ volunteer computing project and it is followed by two follow-up stages on candidates above threshold. We refer to these as Stage-0, Stage-1 and Stage-2. The setup of the \EatH\ search and of the O1 follow-up stages is described in \cite{Ming:2019xse}. The thresholds used here are lower than the ones used in \cite{Ming:2019xse} and candidates exist that survive all O1 follow-up stages. These surviving candidates are further inspected using O2 data.

\subsection{The O1 searches}
\label{subsec:o1stage0}

The searched waveforms are defined by frequency $f$ and frequency derivative parameters $\fdot$ and $\fddot$  in these ranges:
\begin{equation}
\label{eq:Priors}
	\begin{cases}
	-f/ \tau\, \le   \dot{f} ~\le 0\,~\mathrm{Hz/s}\\
	0\,\mathrm{Hz/s}^2 \leq  \ddot{f} \leq ~5\dot{|f|}_{\textrm{max}}^2/f = 5 {f/\tau^2}.
	\end{cases}
\end{equation}
The ages $\tau$ of the compact objects are taken equal to $\ageCasA$ yrs, $\ageVela$ yrs, $\ageG$ yrs for \Cshort, \Vshort, and \Gshort\ respectively. These age values incorporate the uncertainties on the birth-dates of our targets \citep{1997AA...318L..59W,Iyudin:1999qep,Fesen_2006,Allen:2014yra}, airing towards the smaller estimates, which yield a broader (and hence more conservative, safer) search range -- see \cite{Ming:2017anf} for more details.
The frequency derivative ranges for $f=100$ Hz are shown in Table~\ref{tab:SearchParams}, for reference. The frequency and frequency derivatives are defined at the barycentric reference time (epoch of the ephemeris) $\Tref$, which is first given in Table~\ref{tab:SearchParams}.

The basic building block of the hierarchical search is the stack-slide semi-coherent search that uses the global correlation transform (GCT) method \citep{PletschAllen,Pletsch:2008,Pletsch:2010}: The data is divided in segments, and each segment is searched coherently combining the data from both detectors with a matched filter called the $\F$-statisitic \citep{Cutler:2005hc}. The $\F$-statisitic optimally combines together the data from the two detectors at different times based on their noise level and the antenna sensitivity pattern and also references arrival time to the solar system barycenter based on the sub-arcsec coordinates for each target. The results from these coherent searches are appropriately summed, one per segment, yielding an average detection statistic value over all the segments, $\avF$. For a stack-slide search on Gaussian noise, $\Nseg\times 2\avF$ follows a chi-squared distribution with $4\Nseg$ degrees of freedom, $\chi^2_{4\Nseg}$. When $\Nseg=1$ the search becomes a fully coherent search. 

In Table~\ref{tab:GridSpacings} we show the main parameters of the Stage-0 searches: the frequency spacing $\delta f$ and the first and second order frequency derivative spacings $\delta\fdot = \gamma_1 \delta\fdot_c$ and $\delta\fddot = \gamma_2 \delta\fddot_c$. We refer the reader to \cite{Ming:2019xse} for a detailed description of the search setup.

Based on the coherent $\F$-statistics, a new statistics is constructed that is robust to the presence of persistent and transient lines in the data: $\BSNtsc$. This detection statistic tests the signal hypothesis not only against the Gaussian noise hypothesis but also against the hypothesis that a persistent or transient spectral artefact in one of the two data streams is contaminating the data \citep{Keitel:2013wga,Keitel:2015ova}. $\BSNtsc$ eases the impact on the detection efficiency and false alarm rate of various types of spectral artefacts in the data  and we use it to rank the search results on the volunteer computers and based on it, select the results that are returned to the main server. 

A result, or candidate, is defined by the targeted sky position (one of three), by the frequency, first- and second-order frequency derivatives and by its detection statistic value.

We consider only the ``undisturbed'' frequency bands identified in \cite{Ming:2019xse}, i.e. frequency bands where the {\it bulk} of the search results do not display major deviations from what is expected in the absence of large noise artefacts. We apply to the candidates from these bands the same semi-coherent DM-off veto \citep{Zhu:2017ujz} and the same clustering procedure as in \cite{Ming:2019xse,Singh:2017kss}. The DM-off veto discards candidates whose detection statistic increases when the astrophysical Doppler modulation is removed from the searched signal waveform ({\bf{D}}oppler {\bf{M}}odulation-{\bf{off}}). The clustering procedure bundles together high detection statistic values which are close enough in parameter space that they can be ascribed to the same root cause. This saves computations in the follow-up stages because not all close-by candidates are searched independently.

\begin{figure*}
\centering
    \includegraphics[width=0.7 \textwidth]{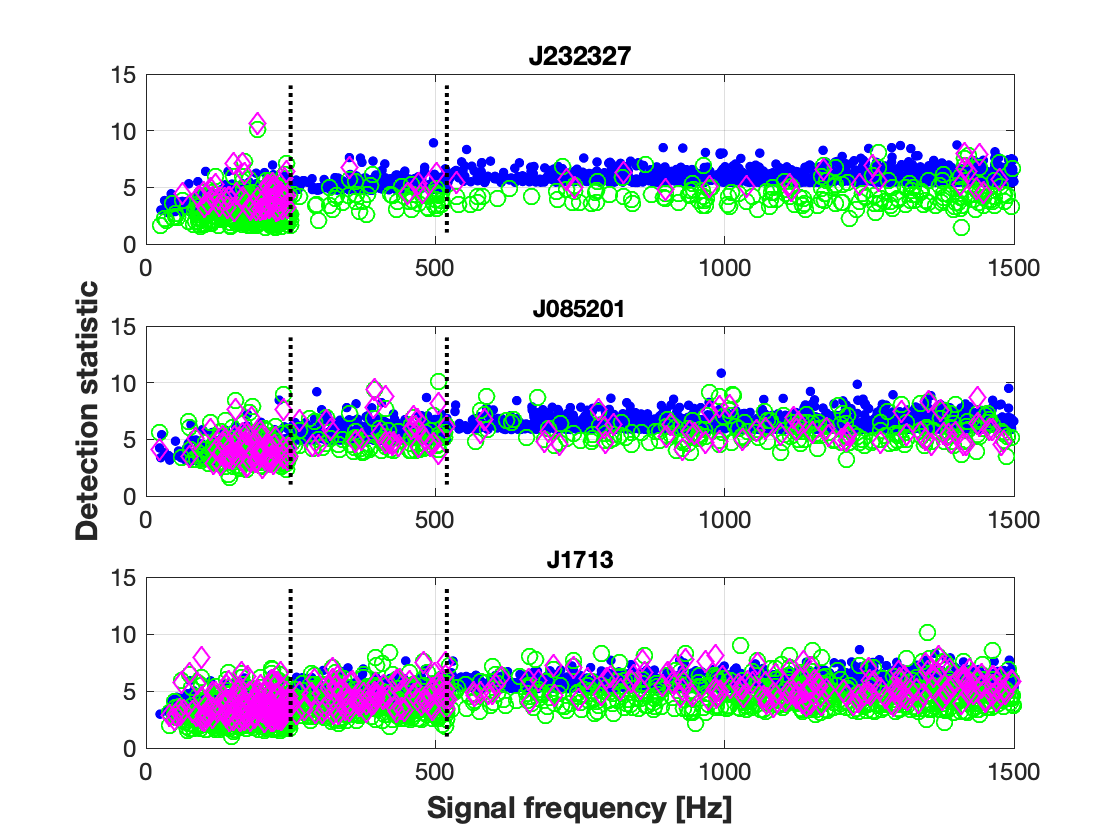}
    \caption{Detection statistic $\BSGLtLr$  as a function of signal frequency of candidates. The smaller filled circles (blue) are the candidates above the Stage-0 threshold that also passed the semi-coherent DM-off veto and are hence searched in Stage-1. The larger (green) circles indicate those candidates that survive Stage-1 and are searched with Stage-2, the fully coherent search using O1 data. The diamonds (magenta) show the candidates that survive  Stage-2. The dotted vertical lines indicate the boundaries of the different frequency ranges for the outlier selection (see Table~\ref{tab:Thresholds}). The maximum value of the detection statistics increases with frequency because the number of searched templates increases with frequency. A J232327 candidate with detection statistic $\sim$ 10 catches the eye, which however is not confirmed in the first follow-up on O2 data (see the middle panel of Figure \ref{fig:O2velacasa}), which means that whatever in the data matched a signal waveform in O1 data yielding that outlier was not there any more in the first half of the O2 data (O2a).}
\label{fig:candidates}
    \includegraphics[width=0.75 \textwidth]{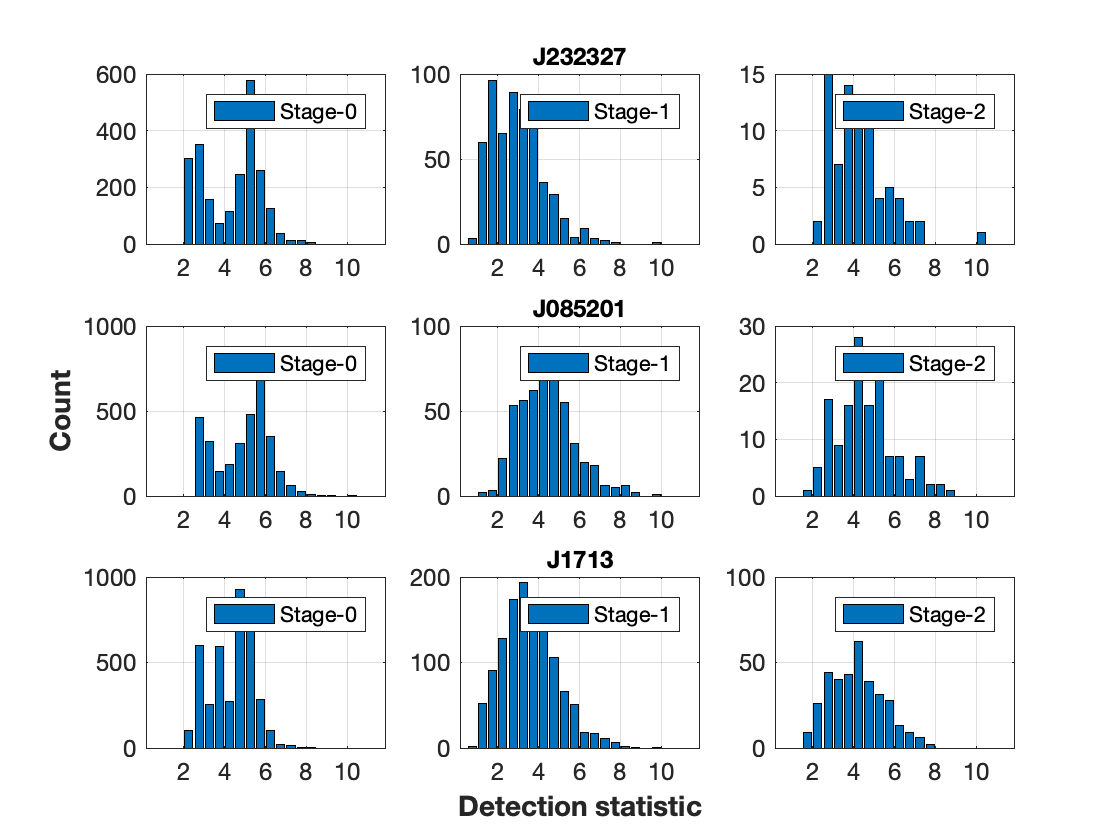}
    \caption{The distributions of the detection statistic $\BSGLtLr$ of candidates that survive each of the three O1 stages, across the entire frequency band. The \Gshort\ candidate at $\approx$ 369 Hz discussed in Section \ref{sec:G347candidate} has $\BSGLtLr \approx 4$ throughout these stages. At this stage the candidate does not stand out among the other outliers under investigation but we point it out in light of the results of the O2a data follow-up.}
\label{fig:CandsStage0}
\end{figure*}

From this set we further select those candidates whose detection statistic value lies above a given threshold. Since the number of templates covered in a fixed frequency interval increases with frequency, we set a different detection threshold  in the frequency ranges  [20-250] Hz, [250-520] Hz and [520-1500] Hz. We refer to these as the low, mid and high frequency range, respectively. Table \ref{tab:Thresholds} shows the values of the thresholds for the different targets and frequencies. The thresholds are such that overall a few thousand candidates are followed up from each target. We use different thresholds for the different targets in the same frequency range because a different number of waveforms are probed for the different targets, due to the different age: for the younger targets a broader range of frequency derivatives is sampled.

The Stage-1 and Stage-2 follow-ups use the same search setup for all targets, with  a coherent time-baseline of about 60 days (1440 hrs) and the entire O1 data set, respectively. Candidates survive from one stage to the next if their detection statistic value increases consistently with what is expected from a signal. 

Initially we have \NcandVelaInStageOne\ candidates from the \Vshort\ search, \NcandCasAInStageOne\ from \Cshort\  and \NcandGInStageOne\ from \Gshort. The DM-off veto rejects \NrejectDMoffVela, \NrejectDMoffCasA\ and \NrejectDMoffG\ candidates for \Vshort, \Cshort\ and \Gshort, respectively. Figure \ref{fig:candidates} shows the detection statistic values of these candidates as a function of frequency throughout these follow-up stages and Figure \ref{fig:CandsStage0} the distribution of their detection statistic values. 

After each follow-up search the number of surviving candidates decreases and at the end of the last stage we are left with $575$ candidates
of which \NcandVelaOutStageTwo\ are from \Vshort, \NcandCasAOutStageTwo\ from \Cshort\ and \NcandGOutStageTwo\ from \Gshort. We note that the fraction of candidates surviving the last stage on O1 data is about 4\% of the original ones for \Vshort\ and \Cshort\ and 9\% for \Gshort. This is consistent with the initial search setup for \Gshort\ having a significantly longer coherence time-baseline compared to the other searches, and hence the follow-up searches having less discriminatory power for the \Gshort\ candidates, that were selected based on more stringent criteria, to start with.

\begin{deluxetable}{cccc}[t]
\tablecaption{Stage-0 detection statistic thresholds. \label{tab:Thresholds}}
\tablehead{
\colhead{\TBstrut} & \colhead{\Cshort} & \colhead{\Vshort } & \colhead{\Gshort }
} 
\startdata
 \TBstrut$~\BSGLtLr^{\texttt{low-freq}}~$ &  $\ThrCasAlow$ & $\ThrVelalow$  & $\ThrGlow$ \\
\TBstrut$~\BSGLtLr^{\texttt{mid-freq}}~$ &  $\ThrCasAmid$ & $\ThrVelamid$  & $\ThrGmid$ \\
\TBstrut$~\BSGLtLr^{\texttt{high-freq}}~$ &  $\ThrCasAhigh$ & $\ThrVelahigh$  & $\ThrGhigh$ \\
\enddata
\end{deluxetable}

\begin{figure}[h!]
\centering
    \includegraphics[width=0.4\textwidth]{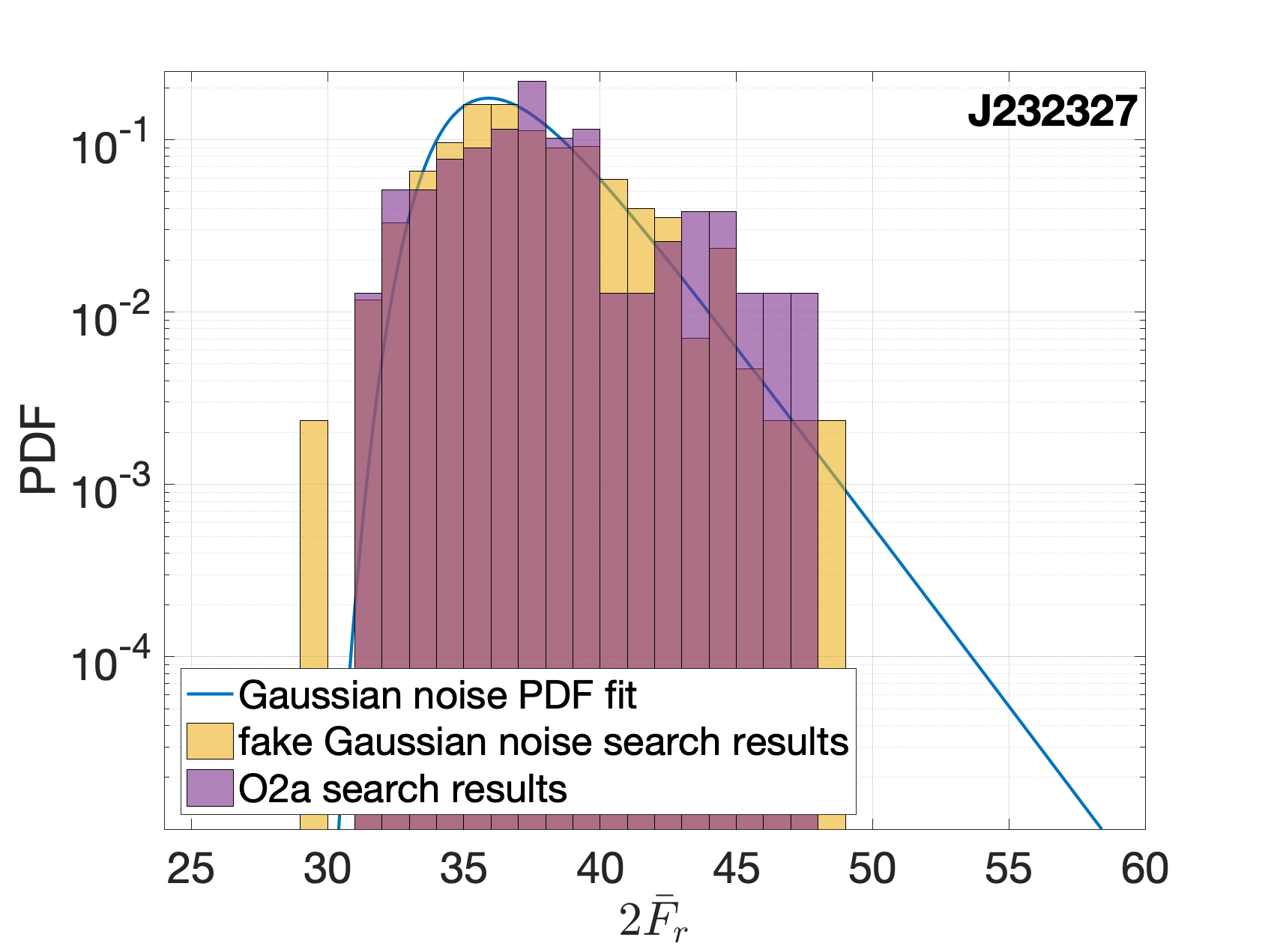}
   \includegraphics[width=0.4 \textwidth]{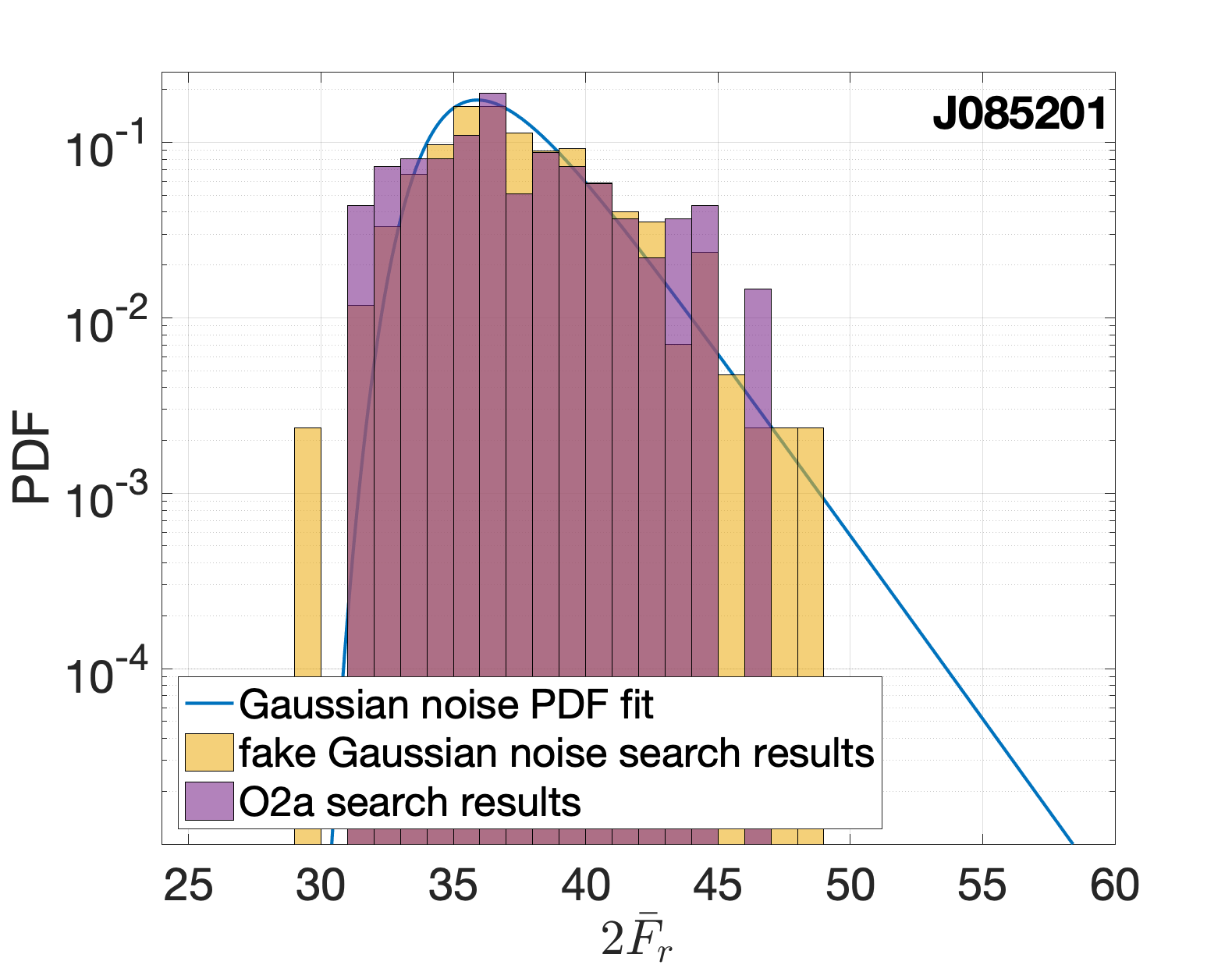}
    \includegraphics[width=0.4 \textwidth]{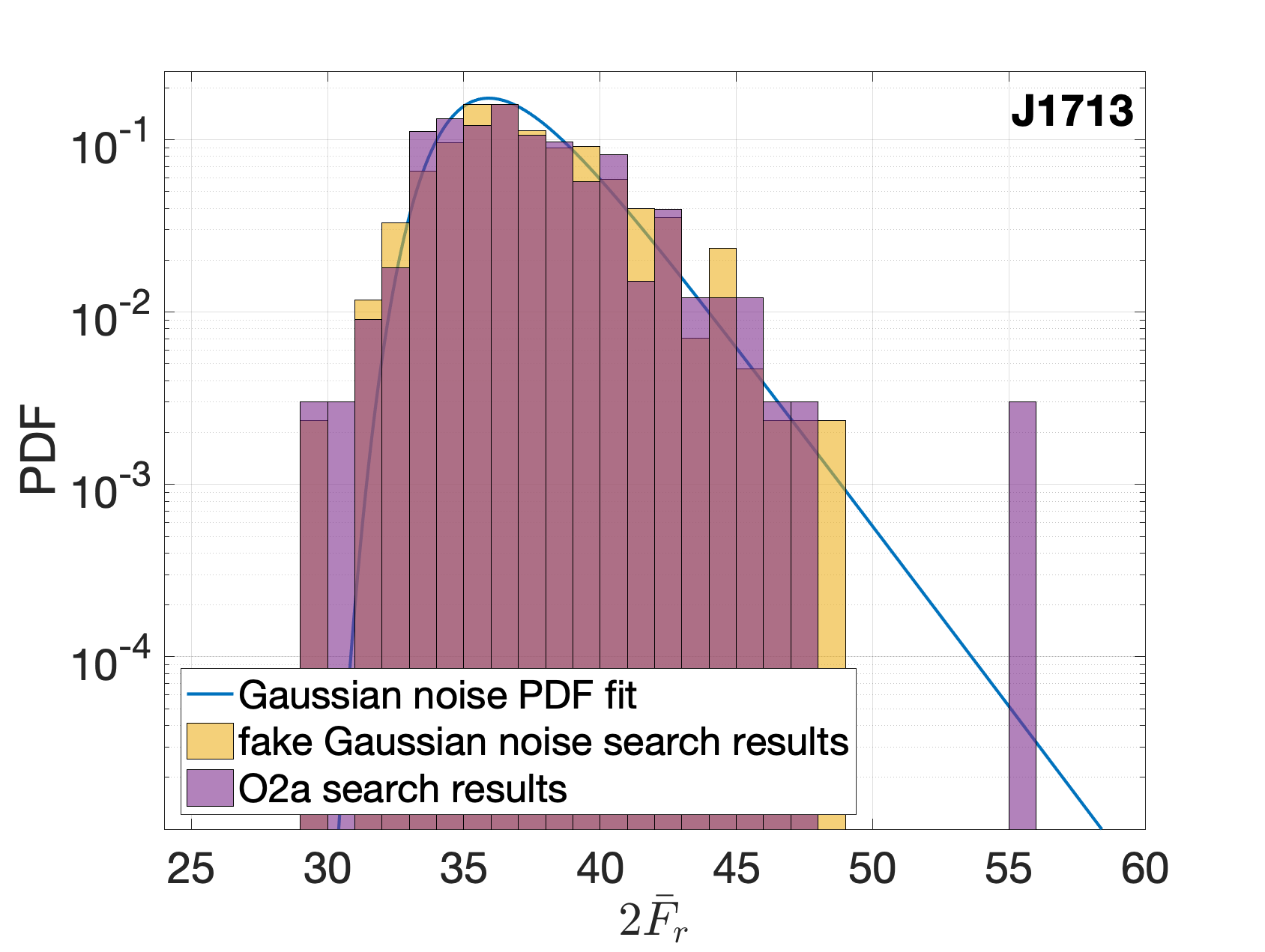}
    \caption{Distribution of the detection statistic $2{\bar\F}_r$ from the O2a searches on each of the \NcandVelaInStageThree~ candidates from \Vshort, \NcandCasAInStageThree~ candidates from \Cshort\ and \NcandGInStageThree~ candidates from \Gshort. The subscript ``$r$" in $2{\bar\F}_r$ refers to the detection statistic being {\bf{r}}ecalculated exactly at the template points. This is a technical detail that we give here for completeness and to explain the subscript.}
\label{fig:O2velacasa}
\end{figure}

\subsection{The follow-up searches on the first half of O2}
\label{subsec:o21search}

In order to reduce the trials factor of the last stage follow-up, we perform two separate searches in O2.  For the first follow-up search on O2 data we only use a portion of the available data, O2a, which is comparable in duration with the O1 set and spans about 120 days. In fact, an interruption in the science run of LIGO between May 9th and June 8th 2017, provides a natural split of the data in two sets. We use the same search setup as for the last stage on O1, which is fully coherent. We search a region which corresponds to the 99\% confidence signal-parameters containment region, based on signal simulation-and-recovery Monte Carlos. We consider the most significant result from each follow-up. 

We follow-up only the outliers that do not fall in O2 spectral regions possibly contaminated by lines: \NcandVelaInStageThree\ from \Vshort, \NcandCasAInStageThree\ from \Cshort\ and \NcandGInStageThree\ from \Gshort, which amounts to $\sim$ 95\% of the O1 outliers. Each search covers about 14 million templates. If these templates had no overlap, the 14 million results would be independent of each other and the distribution of the maximum over every 14 million template search for Gaussian noise would be known. 
In reality nearby templates present a large overlap: the grids are constructed in this way in order to reduce the false dismissal rate for signals whose parameters lie in-between the grid points. We find that the actual distribution for Gaussian noise data is well modelled assuming a smaller number of independent searches (8.2 million). This is what the solid line (red) in Figure~\ref{fig:O2velacasa} shows. The fit was performed on the results of identical searches on fake Gaussian noise, which is what the yellow bars show.

The darker (purple) bars in Figure \ref{fig:O2velacasa} are the normalised histogram for the detection statistic of the results of the searches. For \Cshort\ and \Vshort\ the distributions are completely consistent with Gaussian noise (the lighter, yellow bars). For \Gshort\ an outlier is noticeable at $2\F_r \sim$ 56.  

The p-value associated with the \Gshort\ most significant candidate is about $\pvalueOTwoOneWOneTempl$. This is an estimate of the probability that noise alone would produce a value of the detection statistic like the one associated with that candidate in a single follow-up search. Since \NcandTotInStageThree\ such searches were performed, this translates into an overall p-value of about $4\%$. This is a marginal candidate but we have more data (the second half of O2) that can be used to validate it or discard it. We report further analyses on this candidate in the next section.

The O2a data set comprises \NsftOOneToOTwoOne\% of the amount of data of the O1 data set, hence the detection statistic values from O1 and O2a are remarkably consistent under the assumption of the O1 result being drawn from a distribution with a non-centrality parameter (SNR$^2$) $\rho^2=\TwoFOutlierHR - 4$ and the O2a result being drawn from a distribution with a non centrality parameter  $54=$ \NsftOOneToOTwoOne\% $\rho^2$.

\subsection{The \Gshort\ candidate}
\label{sec:G347candidate}

\subsubsection{The follow-up search on the second part of O2}
\label{sec:g347CandFUs}

We perform a fully coherent search on the second ``half" of O2 (O2b). The amount of data in this set is about 
\NsftOOneToOTwoTwo \% of the O1 data set and \NsftOTwoTwoToOTwoOne \% of the O2a data. The $2\F$ expected in O2b based on O1 (O2a) is 
52.8 (50.1) with a standard deviation of at least 14\footnote{This holds under the assumption that the non-centrality parameter estimate carries no uncertainty. This is not the case, so these numbers are a lower estimate of the standard deviation, which can actually be larger by up to a factor of 2 \citep{prix:2017_coherent_2F_followup}.}.  The expected value for the O2b search due to Gaussian noise (based on Monte Carlos) is less than $\TwoFGNExpectedOTwoTwo$ with a standard deviation of $\sim 2.7$. The measured value of $\TwoFOutlierOTwoTwo$ has a Gaussian p-value of $\TwoFOTwoTwopvalue$ and is within a $1.3$ standard deviations of the expected value on the signal distribution. 

We combine all the data and search the O1+O2 data set. It would be very demanding to carry out this search many times on Gaussian noise in order to provide an estimate of the p-value as done for the O2b search. We hence resort to a conservative estimate of the Gaussian-noise significance based on the number of searched templates. The search covers $\NtemplOOneOTwo$ templates and if they were all independent the maximum expected value for $2\F$ in Gaussian noise would be $\sim 56$ with a standard deviation of $2.7$. The value that we find is  $\TwoFOutlierOOneOTwo$. 
Based on the O1 value, the expected O1+O2 $2\F$ is about $\SignalExpectedOOneOTwo$, with a standard deviation of at least $\SignalStdOOneOTwo$. 
The value that we find does not seem consistent with Gaussian noise but also falls short of the expectations based on the O1 signal-hypothesis by a few standard deviations. 

A summary of the \Gshort\ follow-up results is given in Table \ref{tab:candParams}. 


\begin{deluxetable*}{lccccc}
\tablecaption{Parameters of the most significant results from the high-resolution follow-ups around the most significant candidate from this search, which comes from \Gshort. The barycentric reference time (the epoch of the ephemeris) is $\Tref=\TrefGPS$ GPS s, corresponding to MJD 57345.1986016667 (TDB). \label{tab:candParams}
}
\tablehead{
 $~$   &  \colhead{O1} & \colhead{O2a}& \colhead{O2b} & \colhead{O2} & \colhead{O1+O2} \\
 $~$   &  \colhead{\footnotesize{Sept 2015 - Jan 2016}} & \colhead{Jan 2017 - May 2017}& \colhead{Jun 2017 - Aug 2017} & \colhead{Jan 2017 - Aug 2017} & \colhead{Sept 2015 - Aug 2017} 
    }
\startdata
\TBstrut ~$\Freq$ [Hz] 	&  $\FreqOutlierHR $	& $\FreqOutlierOTwoOne$	& $\FreqOutlierOTwoTwo$	& $\FreqOutlierOTwo$ &  $\FreqOutlierOOneOTwo$  \\
\TBstrut ~$\fdot$	[Hz/s] & $\fdotOutlierHR  $ 	& $\fdotOutlierOTwoOne$ 	& $\fdotOutlierOTwoTwo$		& $\fdotOutlierOTwo$		& $\TBstrut\fdotOutlierOOneOTwo$ \\
\TBstrut ~$\fddot$ 	[Hz/s$^2$] & $\fddotOutlierHR  $ 	& $\fddotOutlierOTwoOne$ 	& $\fddotOutlierOTwoTwo$ 	& $\fddotOutlierOTwo$ 	& $\fddotOutlierOOneOTwo$ \\
\TBstrut ~$2\F$ 	& $\TwoFOutlierHR$ 				& $\TwoFOutlierOTwoOne$ 	& $\TwoFOutlierOTwoTwo$ 	& $\TwoFOutlierOTwo$  & $\TwoFOutlierOOneOTwo$ \\
\TBstrut ~$\propto {\textrm{SNR}}^2 / {\textrm{T}_{data}} $ [s$^{-1}$]	& $\TwoFOutlierHRNormScore$ 				& $\TwoFOutlierOTwoOneNormScore$ 	& $\TwoFOutlierOTwoTwoNormScore$ 	& $\TwoFOutlierOTwoNormScore$  & $\TwoFOutlierOOneOTwoNormScore$ \\
\enddata
\end{deluxetable*}

\subsubsection{X-ray data search}
\label{sec:g347CandEM}

As the central compact objects associated with the targeted SNRs have only been detected in the X-ray energy band, we use archival
X-ray observations of \G\ acquired with sufficient time
resolution to cross-check for an electromagnetic counterpart to the gravitational wave candidate.
 Table~\ref{xraylog} presents the
{\it ROSAT}, {\it Rossi-XTE}, and {\it XMM-Newton} observations used
in this work. These observations are processed, reduced, and analyzed
following the standard procedures for each mission. The photon arrival
times are corrected to the solar system barycenter using the {\it{Chandra}}
coordinates of \Gshort ~\citep{Mignani:2008ew}.

\begin{deluxetable*}{cccccccc}
\tablecaption{Logs of the X-ray observations of RX J1713.4$-$3949 used in our search for an 
electromagnetic counterpart to the \Gshort\ gravitational wave outlier. \label{xraylog}
}
\tablehead{
 \colhead{ObsID} & \colhead{Mission/} & \colhead{Data Mode} & \colhead{Start Date} & \colhead{Exposure} & \colhead{Span} & \colhead{Energy} & \colhead{Res.}\\
 \colhead{} & \colhead{Instr} & \colhead{} & \colhead{UTC} & \colhead{(ks)} & \colhead{(ks)}& \colhead{(keV)} & \colhead{($\mu$s)} 
 }
\startdata 
 500466 & ROSAT/HRI &Nominal     &1996-08-31 & 31  & 901 &0.1-2.5 &   61 \\
  40145-01& RXTE/PCA  &GoodXenon   &1999-06-14 & 8.4 & 905 &   2-5  &    1 \\
0722190101 &XMM/MOS   &CC-mode     &2013-08-24 & 139 & 139 &  0.5-5 & 1500\\
\enddata
\end{deluxetable*}

For the {\it ROSAT} data, source photons are extracted from a
$r<0.2^{\prime}$ circular aperture yielding 1343 counts for the 31~ks
exposure in the 0.1$-$2.5 keV {\it ROSAT} bandpass. The {\it XMM-Newton}
EPIC-MOS data are obtained in one-dimensional {\tt continuous
clocking} mode (CC-mode) with 1.5~ms timing resolution, corresponding to a 
Nyquist frequency of 333.3~Hz. Data are extracted from
the central $0.6^{\prime}$ wide region of the CCD for a total of 59,781
source counts in the $0.5-5$~keV optimal range for the central compact object. The {\it Rossi-XTE}
data is restricted to the top layer of the proportional counter
(2$-$5~keV) whose lower limit just covers the harder end of the soft
central compact object spectrum ($\sim$0.5$-$5~keV), resulting in reduced sensitivity for
the pulsar search using a total of 87,305 counts.

We search these data sets for a coherent pulsed signal over the full Nyquist range including 
the three basic electromagnetic rotation frequency $f_r$ ranges compatible with the
gravitational wave frequency $f$, namely at $f_r=f$, $f_r=f/2$, and the broad r-mode interval
$f_r\approx 3/4 f$.  
We first
look for a strong signal within the full range of possible
frequencies, namely, $100 < f_r < 400$ ~Hz\footnote{With the exception of the {\it XMM-Newton} data set  for which $100 < f_r < 333.3$ ~Hz.}, that allows for the
large time span between the X-ray and LIGO data sets, and the
extrapolated uncertainties in the spin-down measurement,
conservatively estimated as $\delta f=5\times 10^{-5}$~Hz and $\delta
{\dot f}=0.002\times10^{-9}$~Hz/s.  We also conduct a more sensitive
narrow range search around the candidate parameters ($f,
\dot f$) extrapolated to the X-ray data epoch, also allowing for the
uncertainties quoted above.

We use the O1+O2 search parameters given in the last column of Table~\ref{tab:candParams}. The
rapid spin-down associated with this signal requires that we include the first
derivative in the search.  We evaluate the signal significance using an FFT
algorithm, the $Z^2_n$ test \citep{Buccheri:1983zz},
where n=1,2,3 harmonics, and the H-test \citep{dejager1989}, to check
for power at higher harmonics. In these searches, we oversample the
parameter space by a factor of 2.

Given the large time span of the data sets and the small number of
photons, it is most practical to perform the timing search using GPU
technology applied to discrete Fourier transform algorithms ($Z^2_n$, H-test)
instead of FFTs. In the case of {\it XMM-Newton}, the upper range of
the search is limited by the Nyquist frequency for this data set.  For
the {\it Rossi-XTE} data, because of the large span between closely
grouped data segments,  in order to evaluate the power as a
function of frequency, we average fixed sized FFTs on 5 available data
intervals corresponding to these groups.

None of the above searches of the X-ray data yields a significant
($>4 \sigma$) signal close to the expected X-ray frequencies $f_{r}$ and $2f_{r}$, or ones consistent with the r-mode range.

\section{Upper limits}
\label{sec:upperlimits}

\begin{figure}[h!]
\includegraphics[width=0.5\textwidth]{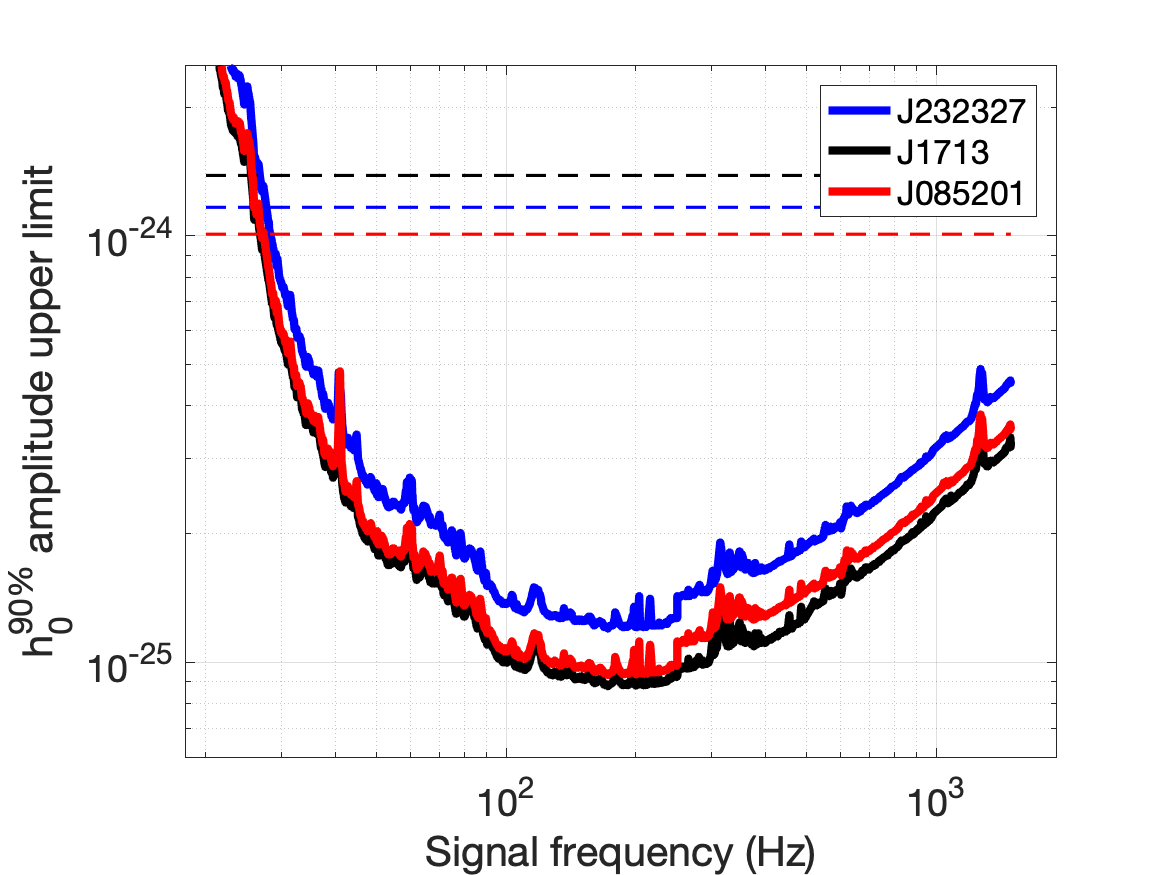}
\includegraphics[width=0.5\textwidth]{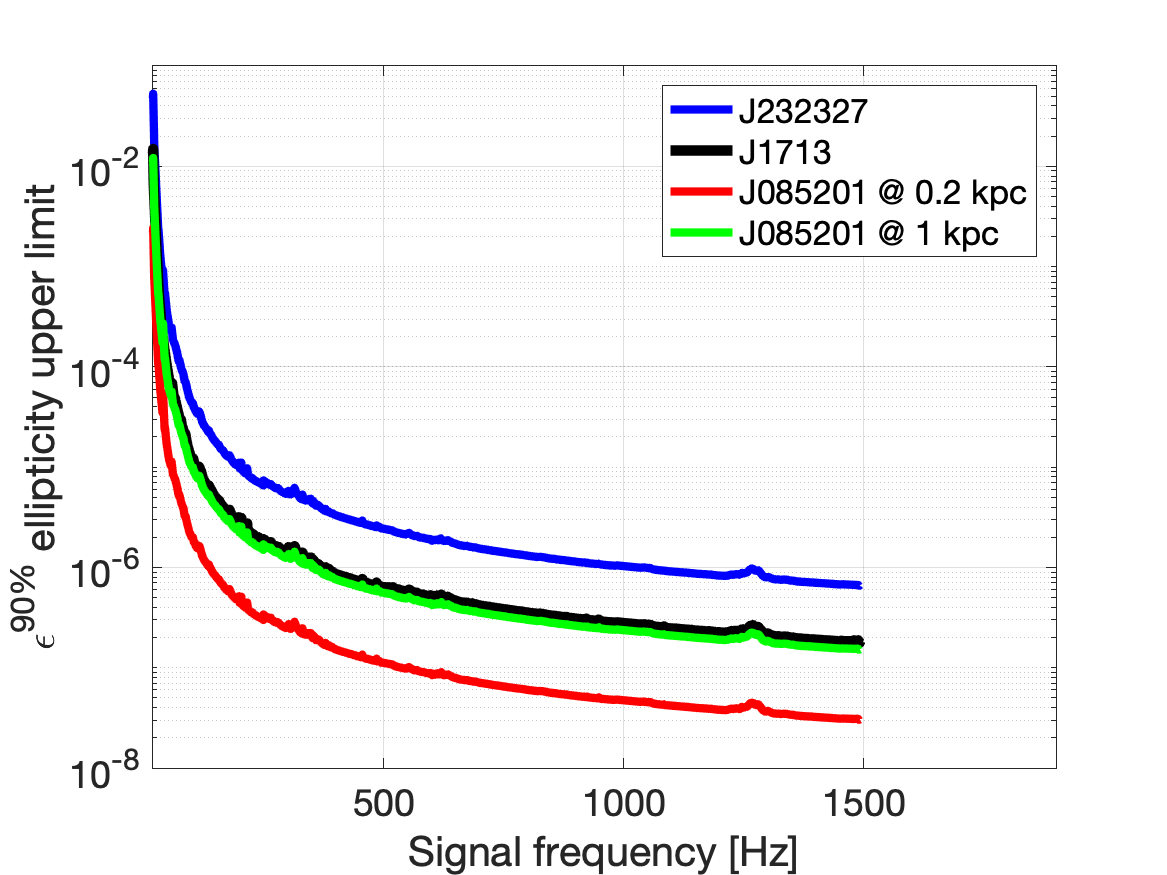}
\includegraphics[width=0.5\textwidth]{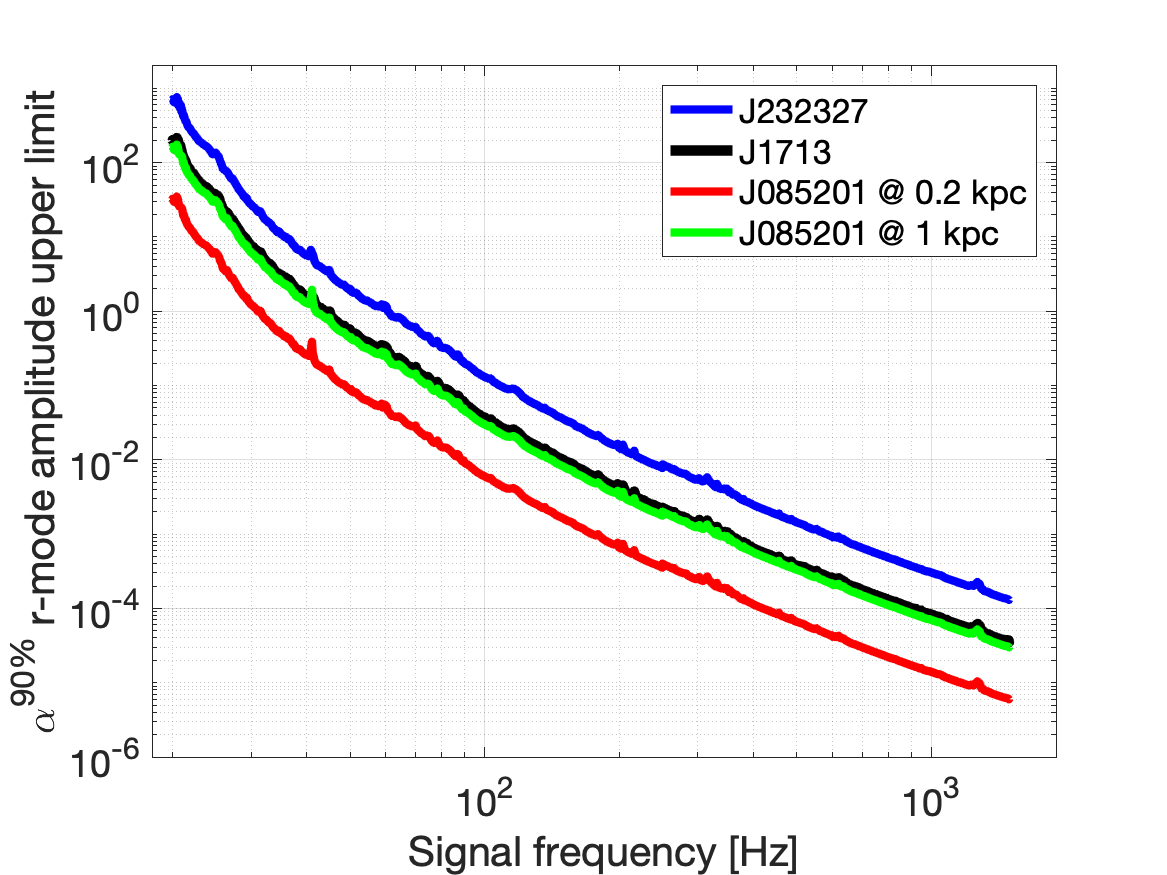}
\caption{90\% confidence upper limits on the gravitational wave amplitude, ellipticity and r-mode amplitude of continuous gravitational wave signals for the three targets. We assume \Vshort\ is either at \distanceVnear\ kpc or at \distanceVfar\ kpc, reflecting the large uncertainty on the actual distance of this object; for \Cshort\ we assume a distance of \distanceC\ kpc and for \Gshort\ a distance of \distanceG\ kpc (see text). The dashed lines in the top panel represent the age-based upper limits for the three targets. For the \Vshort\ line we have assumed a distance of \distanceVnear\ kpc.}
\label{fig:ULs}
\end{figure}

We set frequentist 90\% confidence upper limits on the maximum gravitational wave amplitude consistent with our results as function of the signal frequency,  $h_0^{90\%}(f)$. Specifically, $h_0^{90\%}(f)$ is the expected gravitational-wave intrinsic amplitude such that 90\% of a population of signals with parameter values in our search range would have been detected by our search. In the absence of a detection these represent the smallest amplitude signals that we can exclude. The $h_0^{90\%}(f)$ are determined through search-and-recovery Monte Carlos aimed at measuring the detection efficiency on a set of simulated signals added to the real data. More details of the standard procedure are provided in \cite{Ming:2019xse}. The basic idea is that the confidence level -- 90\% in this case -- is the detection efficiency measured through the Monte Carlos and the $h_0^{90\%}(f)$ corresponds to the signal amplitude such that the detection confidence is 90\%. The most constraining upper limits are all close to signal frequencies of $\sim$ \lowestULvfreq\ Hz and are \lowestULc\ for CasA, \lowestULv\ for \Vshort\ and \lowestULg\ for \Gshort. Figures~\ref{fig:ULs} show all the upper limit curves discussed in this section.

The sensitivity depth ${{\mathcal{D}}}^{90\%}$ \citep{Behnke:2014tma, Dreissigacker:2018afk} expresses the sensitivity of a search: 
\begin{equation}
{{\mathcal{D}}}^{90\%}:={\sqrt{S_h(f)}\over {h_0^{90\%}(f) }}~~[ {1/\sqrt{\text{Hz}}} ],
\label{eq:sensDepth}
\end{equation}
where $\sqrt{S_h(f)}$ is the noise level associated with a signal of frequency $f$, it is the same as the one reported in \cite{Ming:2019xse} and it is available in machine-readable format at \cite{aeiurl}. Table \ref{tab:sensDepth} shows the sensitivity depth values of the searches in the different frequency ranges. As expected, the sensitivity depth is larger for the targets searched with an initial longer coherent baseline, i.e. for the older targets (\Gshort\ is the eldest and then comes \Vshort). Furthermore, as the number of spindown templates increases, due to the trials factor, the detection threshold is increased (see Table~\ref{tab:Thresholds}) and the search becomes less sensitive. For this reason 
or the same target, the sensitivity depth is higher at lower frequencies. 

\begin{deluxetable}{l|ccc}
\tablecaption{Sensitivity depth $~{{\mathcal{D}}}^{90\%}$ from Eq.~(\ref{eq:sensDepth}) corresponding to the $h_0^{90\%}$ upper limits set by this search. The numbers in parenthesis are the sensitivity depths for the higher threshold search described in \cite{Ming:2019xse}.\label{tab:sensDepth}
\label{tab:sensDepth} 
}
\tablehead{
& \multicolumn{3} {c}{Sensitivity depth ${{\mathcal{D}}}^{90\%} ~[{1/\sqrt{\text{Hz}}}] $} \\
{Freq. range}  & \colhead{\Cshort} & \colhead{\Vshort } & \colhead{\Gshort}
}
\startdata
20-250   Hz & 65.4 (60.3) &  84.2 (76.5) & 89.4 (82.9) \\
250-520 Hz & 58.0 (54.5) &  74.3 (70.1) & 85.6 (79.1)\\
520-1500 Hz & 57.8 (53.7) & 73.9 (70.0) & 81.4 (76.4) \\
\enddata
\end{deluxetable}

The upper limits on the amplitude can be readily translated in upper limits on the equatorial deformation $\epsilon$ of the star (assuming the moment of inertia in the direction of the spin axis $10^{38}$ kg m$^2$) and r-mode amplitude $\alpha$ (under the same assumptions as \cite{Owen:2010ng}), as follows:
\begin{equation}
\begin{aligned}
\varepsilon &=2.8 \times 10^{-7} ~\left( {h_0\over{3\times10^{-25}}}\right )\left ( {D\over{1~\textrm{kpc}}}\right ) \left ({{\textrm{1000~Hz}}\over f} \right )^2\\
\alpha &= 8.4 \times 10^{-6}  \left( {h_0\over{3\times10^{-25}}} \right )\left ( {D\over{1~\textrm{kpc}}}\right ) \left ({{\textrm{1000~Hz}}\over f} \right )^3.
\end{aligned}
\label{eq:epsilon}
\end{equation}
For the upper limits on $\varepsilon$ and $\alpha$ shown in Figure \ref{fig:ULs} we set the distances of the targets as follows: for \Cshort\ we take \distanceC\ kpc, following the estimate $3.4^{+ 0.3}_{- 0.1}$ kpc of \cite{Reed:1995zz}; for \Gshort\ we take \distanceG\ kpc which is the center of the interval $1.3{\pm 0.4}$ kpc of \cite{CassamChenai:2004ss}; there is no consensus on the distance of \Vshort\ so, from \cite{Allen:2014yra} we consider two extremes: a distance of \distanceVnear\ kpc and of \distanceVfar\ kpc, corresponding to the nearest and farthest that the object could be.

\section{Conclusions}
\label{sec:conclusions}

We follow-up about 10,000 sub-threshold candidates from an Einstein@Home search on O1 LIGO data for signals from \Vshort, \Cshort, and \Gshort\ - the central compact objects in the supernova remnants Vela Jr., Cassiopeia A and SNR G347.3$-$0.5, respectively. At the threshold values set in this search a confident detection could not be claimed based on O1 data alone, but our follow-up searches on the newly released LIGO O2 data would be able to confirm a candidate stemming from a real signal. 

Only one candidate from the first O2 follow-up warrants further investigation. This is at a frequency of about 369 Hz and comes from the search that targets emission from \Gshort. The follow-up search results on the remainder of O2 data (O2b) fall short of the expected value and overall are not completely conclusive. If this candidate were of astrophysical origin it is unlikely that it would be a perfectly continuous and phase-coherent signal across all the observing periods that we have examined. We note that the search by \cite{Millhouse:2020jlt}, which is robust against timing noise, does not produce any outlier consistent with ours, albeit the sensitivity of that search is not clear. 

The timing solutions of Table~\ref{tab:candParams} for the \Gshort\ outlier imply a characteristic spin-down age of $\sim 1,300$ years, a spin-down energy loss rate of $1.6\times 10^{40}$ erg/s and a surface magnetic field of $6\times 10^{11}$ G. The characteristic age is consistent with that of a young NS associated with a undispersed supernova 
remnant ($\lesssim {\rm few} \times 10^{4}$). The implied magnetic field is higher than of the detected central compact object pulsars 
but at the lower range for the rotation-powered pulsars. The large spin-down energy is in excess of the most energetic pulsars 
known (c.f. Crab pulsar, $\dot E = 4.6\times 10^{38}$~ergs/s) and nearly all pulsars with $\dot E\gtrsim 4\times 10^{36}$~ergs/s display bright X-ray and/or radio pulsar wind nebulae \citep{2004IAUS..218..225G}, with several notable exceptions.  No 
wind nebula are associated with the known central compact object pulsars.

In the spirit of ``leaving no stone unturned'' we conduct an exhaustive timing search to identify an electromagnetic
counterpart for this candidate.  Since this source is
only seen in X-ray emission, we analyze all available archival X-ray
data on \Gshort\ acquired with sufficient timing resolution.  Although we
do not find a significant signal at any of the expected frequencies, the
sensitivity of the search is low, due to the many years
gap between the LIGO and X-ray observations. This gap produces a very high number of search trials that results in a loss of sensitivity of our search. 

New X-ray observations, contemporaneous with the LIGO runs, are planned, which should
provide a more sensitive assessment\footnote{At the same X-ray signal strength the observation will be orders of magnitude more significant.} of the \Gshort\ candidate. A gravitational wave search in O3 data based on the parameters provided here should be easily able to shed light on the nature of this outlier. Noise investigations on the LIGO instruments could also reveal that the origin of outlier is a coherent contamination.

We set new upper limits on the intrinsic continuous gravitational wave amplitude from \Cshort, \Vshort, and \Gshort, improving on those already set in \cite{Ming:2019xse}. Our upper limits are below the age-based indirect upper limit (see for instance section 3 of \cite{Wette:2008hg} for a definition) over most of the searched frequency range. Ellipticities at and above $10^{-6}$ are excluded for all targets above 1000 Hz (rotation frequency 500 Hz) and for \Gshort\ above $\sim$ 370 Hz (rotation frequency 185 Hz). In fact for \Gshort\ our results exclude ellipticities larger than  $10^{-7}$ above 920 Hz. This range of ellipticity values is interesting: neutron star crusts should readily be able to sustain ellipticities of the order of $10^{-6}$ \citep{McDanielJohnsonOwen,Baiko:2018jax}. If r-modes are the gravitational-wave emission mechanism, our upper limits exclude amplitudes greater than $10^{-3}$ for all targets above $\sim$ 580 Hz ($\approx$ 435 Hz rotation frequency) and for \Gshort\ above $\sim$ 360 Hz (rotation frequency $\approx$ 270 Hz). In fact for \Gshort\ our results exclude r-mode amplitudes larger than  $5\times 10^{-4}$ above 440 Hz (rotation frequency $\approx$ 330 Hz).  Estimates of the r-mode amplitude yield values of $10^{-3}$ or higher for several neutron-star hydrodynamic evolution scenarios \citep{Bondarescu:2008qx,Owen:2010ng}. These are well within the reach of this search.

The focus of the work presented here is however not in the upper limits but in detection, i.e. investigating whether any of the $\approx 10,000$ outliers examined was due to a gravitational wave signal. Since the sensitivity of our search remains unmatched \citep{Lindblom:2020rug,Millhouse:2020jlt}, investigating 10,000 new candidates further moves in uncharted territory. 

This work also  presents the first extensive follow-up of a continuous gravitational wave candidate in electromagnetic data and sets the stage for more investigations of this nature, including electromagnetic sub-threshold surveys followed by focused gravitational wave searches. It will be a combination of sensitive data and ingenuous search techniques, perhaps comprising multi-messenger observations, that will eventually secure the first detection of a continuous gravitational wave signal.

\section{Acknowledgments}

The authors gratefully acknowledge the support of the many thousands of Einstein@Home volunteers, without whom this search would not have been possible. 

M.A. Papa and B. Allen gratefully acknowledge support from NSF PHY grant 1816904. 
E.V. Gotthelf  acknowledges support from NASA grant 80NSSC18K0452 \& SAO grant GO9-19046X.

The authors thank the LIGO Scientific Collaboration for access to the data and gratefully acknowledge the support of the United States National Science Foundation (NSF) for the construction and operation of the LIGO Laboratory and Advanced LIGO as well as the Science and Technology Facilities Council (STFC) of the United Kingdom, and the Max-Planck-Society (MPS) for support of the construction of Advanced LIGO. Additional support for Advanced LIGO was provided by the Australian Research Council.
This research has made use of data, software and/or web tools obtained from the LIGO Open Science Center (\url{https://losc.ligo.org}), a service of LIGO Laboratory, the LIGO Scientific Collaboration and the Virgo Collaboration, to which the authors have also contributed. LIGO is funded by the U.S. National Science Foundation. Virgo is funded by the French Centre National de Recherche Scientifique (CNRS), the Italian Istituto Nazionale della Fisica Nucleare (INFN) and the Dutch Nikhef, with contributions by Polish and Hungarian institutes.
The follow-up studies for the \Gshort\ outlier have also employed the {\tt{PyFstat}} software suite \citep{ashton_gregory_2020_3620861} and the Falcon search code \citep{Dergachev:2019wqa}.
\bibliography{paperBibApJ}{}
\bibliographystyle{aasjournal}

\end{document}